\documentclass[letter, traditabstract]{aa}
\usepackage{graphicx}
\usepackage{txfonts}
\usepackage{amssymb}
\usepackage[below]{placeins}
\usepackage{natbib}
\usepackage{lscape}
\usepackage{threeparttable}
\usepackage{sidecap}
\bibpunct{(}{)}{;}{a}{}{,}

\def \rad500{r_{500}}
\def \t500{T_{500}}
\def \m500{M_{500}}

\def \xspt{SPT-CL\ J2332-5358}
\def \xsptt{SPT-CL\ J2342-5411}
\def \xmm{\emph{XMM-Newton}}

\begin{document}

%________________________________________________________________

%\title{X-ray detection of two clusters with strong Sunyaev-Zel'dovich effect signatures in the South Pole
%Telescope Survey}
\title{XMM-Newton detection of two clusters of galaxies with strong SPT Sunyaev-Zel'dovich effect signatures}
\titlerunning{XMM-Newton detection of two clusters with strong SPT SZE signals}
%\subtitle{}

\author{R.~\v{S}uhada\thanks{email: rsuhada@mpe.mpg.de} \inst{1},
J.~Song \inst{2},
H.~B\"ohringer\inst{1},
B.~A.~Benson\inst{3,5,6},
J.~Mohr\inst{1,4,10},
R.~Fassbender\inst{1},
A.~Finoguenov \inst{1,9},
D.~Pierini\inst{1},
G.~W.~Pratt\inst{7},
K.~Andersson\inst{8},
R.~Armstrong\inst{2}  and
S.~Desai\inst{2}
}

\institute{
\inst{1} Max-Planck-Institut f\"ur extraterrestrische Physik, Giessenbachstr. 1, 85748 Garching, Germany \\
\inst{2} University of Illinois, Department of Astronomy, 1002 West Green St, Urbana, IL 61801, USA \\
\inst{3} Kavli Institute for Cosmological Physics, University of Chicago, 5640 South Ellis Avenue, Chicago, IL 60637, USA \\
\inst{4} Department of Physics, Ludwig-Maximilians-Universit\"{a}t, Scheinerstr. 1, 81679 Munich, Germany \\
\inst{5} Department of Physics, University of California, Berkeley, CA 94720, USA \\
\inst{6} Enrico Fermi Institute, University of Chicago, 5640 South Ellis Avenue, Chicago, IL 60637, USA \\
\inst{7} Laboratoire AIM, IRFU/Service d'Astrophysique - CEA/DSM - CNRS - Universit\'{e} Paris Diderot, B\^{a}t. 709, CEA-Saclay, F-91191 Gif-sur-Yvette Cedex, France \\
\inst{8} MKI, Massachusetts Institute of Technology, Cambridge, MA 02139, USA \\
\inst{9} University of Maryland, Baltimore County, 1000 Hilltop Circle, Baltimore, MD 21250, USA \\
\inst{10} Excellence Cluster Universe, Boltzmannstr. 2, 85748 Garching, Germany \\
}
\date{Received/accepted}
%________________________________________________________________

% \abstract{}{}{}{}{}
% 5 {} token are mandatory

  % \abstract
   %{tbd}
   %{tbd}
   %{tbd}
\abstract {We report on the discovery of two galaxy clusters, \xspt\ and \xsptt, in X-rays.
These clusters were also independently detected through their Sunyaev-Zel'dovich effect
by the South Pole Telescope, and confirmed in the optical band by the Blanco Cosmology Survey.
% and in the optical band by the Southern Cosmology Survey
They are thus the first clusters detected under survey conditions by all major cluster search approaches.
The X-ray detection is made within the frame
of the XMM-BCS cluster survey utilizing a novel \xmm\ mosaic mode of observations. The present study makes the first scientific use of this operation mode. 
We estimate the X-ray spectroscopic temperature of \xspt\ (at redshift z=0.32)  to T $=9.3^{+3.3}_{-1.9}$~keV,
implying a high mass, M$_{500} = 8.8 \pm 3.8 \times 10^{14} $~M$_{\sun}$.
For \xsptt,  at z=1.08, the available X-ray data doesn't allow us to directly estimate the temperature with
 good confidence. However, using our measured luminosity and scaling relations we estimate that T$=4.5\pm{1.3}$~keV and M$_{500} = 1.9 \pm 0.8 \times 10^{14} $~M$_{\sun}$.
We find a good agreement between the X-ray masses and those estimated from the Sunyaev-Zel'dovich effect.
}

\authorrunning{R. \v{S}uhada et al.}
\maketitle
%
%________________________________________________________________

\section{Introduction}
\begin{figure*}[t!]
	\begin{minipage}{0.665\textwidth}
	\begin{center}
		\includegraphics[width=1.00\textwidth]{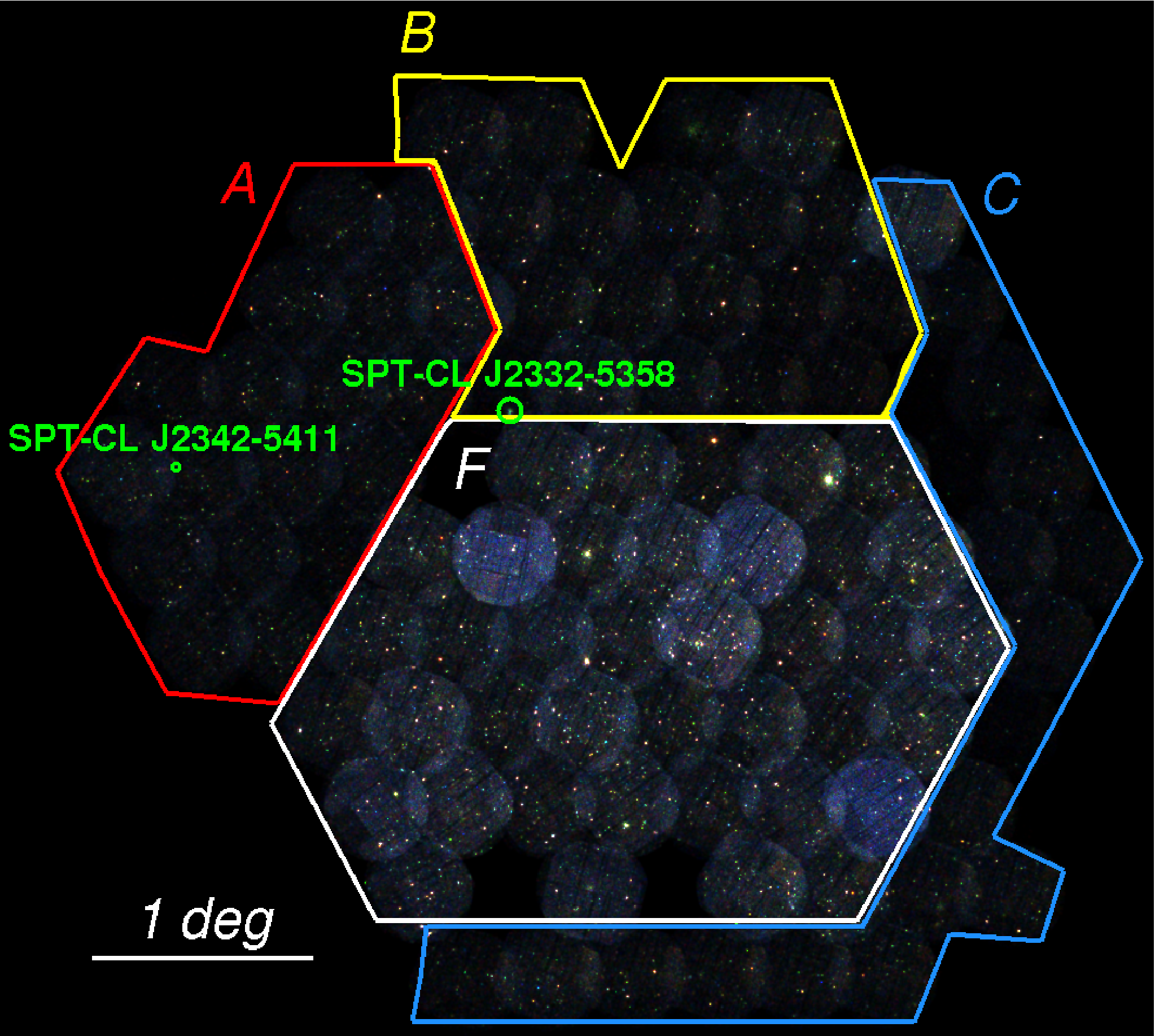}
	\end{center}
	\end{minipage}
	\begin{minipage}{0.335\textwidth}
	\begin{center}
		\includegraphics[width=0.96\textwidth]{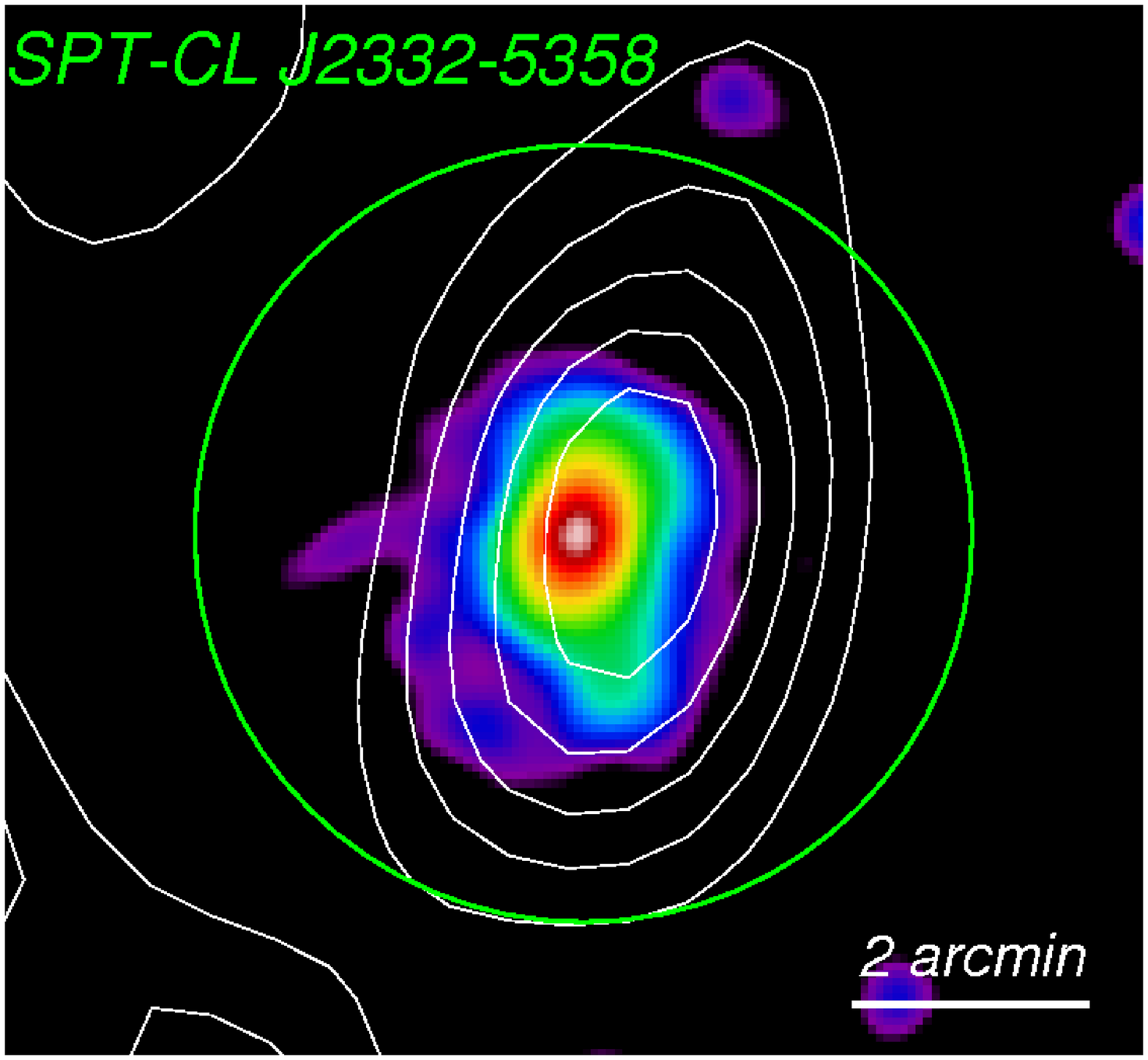}
		\includegraphics[width=0.96\textwidth]{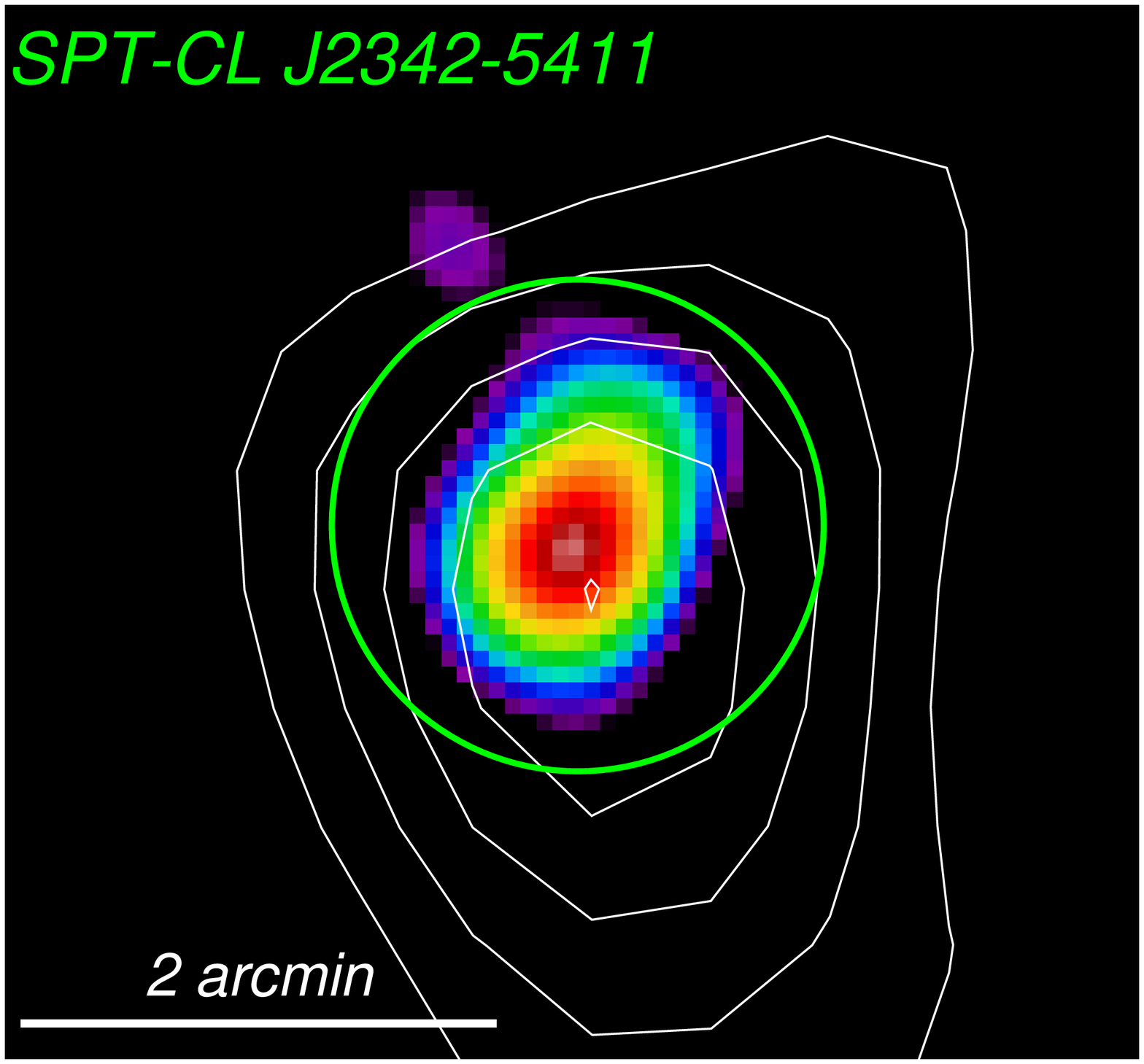}
	\end{center}
	\end{minipage}
	\caption{\emph{Left:} Mosaic mode \xmm\ image of the whole 14 deg$^2$ survey field.
The RGB false color image was constructed from surface brightness images in
 the $0.3-0.5$, $0.5-2.0$ and $2.0-4.5$~keV bands. Regions A, B and C mark the three mosaic mode observations, region F the deeper core of the survey consisting of 42 individual pointings.
 The green circles mark the positions of \xspt\ and \xsptt\ with a radius equal to r$_{\mathrm{plat}}$ (Sec.~\ref{sec:gca}) in both
images. \emph{Right:} $0.5 - 2.0$~keV images of \xspt\ (top) and \xsptt\ (bottom) with overlaid SZE
signal-to-noise contours from the SPT survey (V10).}
\label{fig:mosaic}
\end{figure*}
Almost 40 years after the theoretical prediction of the Sunyaev-Zel'dovich effect \citep[SZE,][]{sze72}, i.e. the distortion
of the cosmic microwave background spectrum by the hot
gas in clusters of galaxies,
 we have entered a new era where the first clusters have now been discovered by
large area SZE surveys \citep{staniszewski09}. Two ambitious SZE
cluster surveys are currently underway: by the South Pole Telescope (SPT)
and by the Atacama Cosmology Telescope (ACT).  Recently, the SPT released a
catalog of 21 SZE selected galaxy clusters identified in the first $\sim200$
deg$^2$ of sky surveyed by the SPT \citep[][hereafter V10]{vanderlinde10}. 
Both SPT and ACT have additionally carried out observations of
known clusters \citep{plagge09, hincks09}.

The SZE provides new prospects for precision cluster cosmology for two main reasons:
\textbf{(1)} the SZE decrement characterized by the Comptonization parameter Y
is currently considered as a robust, low-scatter proxy for cluster mass
\citep[e.g.][]{dasilva04,motl05}
and \textbf{(2)} the SZE is not subject to the cosmological surface brightness
dimming effect, resulting in a selection
function closely corresponding to a selection with a fixed mass limit at all redshifts.

However, to be able to fully harvest the potential of the upcoming comprehensive
multi-wavelength surveys, we need to have \textbf{(1)} a good understanding of
the cluster selection function, \textbf{(2)} cluster redshift measurements and
\textbf{(3)} a well calibrated link between cluster observables and total cluster
masses.

In order to address these issues and to best understand the results of the
different survey techniques, we are conducting a
coordinated multi-wavelength survey in a test region (which will be covered by
both SZE-surveys) in the optical by the Blanco Cosmology Survey (BCS,
100 deg$^2$), in the mid-infrared with \emph{SPITZER} (14 deg$^2$) and in X-rays with \xmm.

Here we present the X-ray detections of two clusters, \xspt\ and \xsptt. These clusters were
 independently detected by SPT (V10) and confirmed to be coincident with overdensities of red galaxies \citep{high10}.
In addition, \xspt\ has been recently detected in the optical \citep[SCSO~J233227-535827,][]{menanteau10}.
 This source is also coincident with the X-ray source 1RXS~J233145.2-534616 in the ROSAT
 faint source catalogue \citep{voges00}, but couldn't be identified as an extended source.
  The present \xmm\ observations enable us to confirm both objects as X-ray luminous clusters of galaxies.

Throughout the article, we adopt a $\Lambda$CDM cosmology with
$(\Omega_{\Lambda}, \Omega_{M}, H_0) = (0.7, 0.3, 70$ km s$^{-1}$ Mpc$^{-1})$.
\section{XMM-Newton data reduction}
\xspt\ and \xsptt\ were discovered as high significance extended sources in \xmm\ 
observations carried out in the framework of the XMM-BCS cluster survey
(\v{S}uhada et al., in prep.).

The X-ray survey currently extends over 14
deg$^2$  (Fig.~\ref{fig:mosaic}).
 The core of the \xmm\ field consists of a deeper region covering 6 deg$^2$ with
42 partially overlapping $\sim 12$ ks long individual pointings and three
large scale ($\sim 2.7$ deg$^2$ each) \emph{mosaic mode observations}.
 Each of the three mosaics consists of 19
stable pointings (3.5 ks exposures) and the slews between them, with a total time
$\sim90$ ks per mosaic.

%can be cut
The mosaic mode extension of the X-ray survey was designed to match in extent the existing
\emph{SPITZER} coverage (i.e. $\sim 14$ deg$^2$), while requiring only shallow depth
sufficient to roughly match the SPT sensitivity determined tentatively after the
start of its operations.

\subsection{XMM-Newton mosaic mode observations}
The \emph{mosaic mode observation} is a new observation mode of \xmm\ 
and this is the first instance of its scientific use. Mosaic mode observations
were designed to significantly increase the efficiency
of observations covering areas larger than the field of view of the telescope.
Before the implementation of this mode such observations could only be achieved
by consecutive independent single pointings. Each of these individual pointings
then required its own instrumental overhead, which particularly for the case of the
EPIC pn camera can be a significant part of the total observing time, especially
if the required exposure times for the pointings themselves are low.

The mosaic mode observation starts as a standard observation with operational
overhead (telescope pointing and guide star acquisition) followed by
instrumental overhead, when a charge zero level (i.e. \emph{offset table}) is
calculated for the pn camera, which amounts typically to $3-4$ ks (MOS cameras
are operated with fixed offset tables and their setup is negligible). After the setups are finished, the observation itself starts. In our mosaics,
each stable pointing has an exposure of 3.5 ks, followed by a slew to the next
field offset by $\sim 23\arcmin$. Science data is collected also during the
slew and at variance with the standard operating mode the observation is not interrupted
by a new instrumental setup sequence, but the same offset table is used during the
whole mosaic.

Without the mosaic mode, surveys of this kind would practically be
unfeasible, with observing efficiency (i.e. the ratio of integration time to
total time) around only $50\%$, compared to $\gtrsim 80\%$ efficiency achieved
with the present setup. More information on the mosaic mode observations can be
found in the \xmm\ User Handbook\footnote{\texttt{
xmm.esac.esa.int/external/xmm\_user\_support\\/documentation/uhb/XMM\_UHB.pdf}} .

\subsection{X-ray data analysis}
\label{sec:gca}
Both \xsptt\ and \xspt\ were detected in the mosaic observations carried out in December 2009
(mosaic A, OBSID: 0604870301 and mosaic B, OBSID: 0604873401 respectively).
We defer a more detailed description of the survey data reduction to a
forthcoming publication of the X-ray cluster catalog.
Here we summarize the main steps and highlight the differences of treating
mosaic eventlists with respect to standard observations.

The EPIC data is processed with the current \emph{XMM-Newton} Standard Analysis
System (SAS) version 9.0.0. We calibrate the raw observational data files in a standard
way. Events in bad pixels, bad columns and close to the chip
gaps are excluded from further analysis.
The eventlists were screened for high background periods caused by soft proton
flares following the two step cleaning method of \citet{pratt03}, but setting a
 $3\sigma$ limit in both %marker%
energy bands.

The clean exposure times are 71.9/72.6 ks for pn, and 85.0/88.7 ks for the MOS
cameras for the entire mosaic A/B respectively. The
beginning of the mosaic sequence B in the pn camera was strongly affected
by soft proton flaring, therefore the effective exposure at the \xspt\ location is only 0.1 ks
in pn, while it is $3.1$ ks in each MOS camera.
The source in addition lies
partially on the missing MOS1 CCD\#6 yielding a total combined MOS effective
exposure of only $\sim 4$ ks.
Local exposure times for \xsptt\ are $\sim 2.8$~ks in pn, $\sim 2.1$~ks in MOS1
 and $\sim 2.3$~ks in MOS2.

As the main source detection algorithm we utilize the sliding box
technique and a  maximum likelihood source fitting in their current, improved
implementation in the SAS tasks \texttt{eboxdetect} and \texttt{emldetect}.

Mosaic data of this extent is too large to fit into the memory storage during the
detection process. Therefore we segment the mosaic into several overlapping parts,
which can be handled by the SAS tasks. Segmenting the mosaic into sky-chunks 
for source detection is preferable to splitting it into individual stable pointings, 
because we also want to include counts gathered during the slews between the
pointings, as well as utilize the greater depth in the regions where two
neighbouring pointings overlap.

The mosaic segments have a typical size of $\sim 1$ deg$^2$ and overlap by
$\gtrsim 2\arcmin$ along all borders. This way the
input images, exposure and background maps can be accommodated by the
\texttt{ebox-} and \texttt{emldetect} tasks ran with increased memory buffer
(imagebuffersize=2000 flag).

In order to get a reliable measurement of the flux and trace the emission of
the clusters as far out as possible, we implemented a refined version of the
\emph{growth curve method} \citep[][\v{S}uhada et al., in prep.]{boehringer00}. The cumulative source flux
as a function of radius (i.e. the growth curves) for the two systems are
displayed in Fig.~\ref{fig:gca1} and Fig.~\ref{fig:gca2} (in the online Appendix).
The total source flux is determined iteratively by fitting a line to the flat part of the background subtracted growth curve.
We define the \emph{plateau radius} (r$_{\mathrm{plat}}$) as the aperture where the growth curve reaches the total flux.

For \xspt, we detect source emission out to r$_{\mathrm{plat}}=196\arcsec$, with total source flux of F$_{\mathrm{plat}}(0.5 - 2.0$~keV$) = 9.38 \pm 0.50 \times 10^{-13}$ erg s$^{-1}$ cm$^{-2}$,
 corresponding to a total luminosity L$_{\mathrm{plat}}(0.5 - 2.0$~keV$) =~2.67~\pm~0.14~\times~10^{44}$~erg~s$^{-1}$. Errors of the flux and luminosity include the Poisson errors and a 
5\% systematic error in the background estimation.

The X-ray morphology of this cluter exhibits a good agreement with the SZE signal on the largest scales (Fig.~\ref{fig:mosaic})
and its peak is close to the position of the brightest cluster galaxy (BCG, Fig.~\ref{fig:opt}).
We detect a significant X-ray extension up to $\sim 1.5\arcmin$ SE from the BCG. Detailed characterization
 of the galaxy distribution and the correlation between X-ray and optical morphology will be addressed in a forthcoming paper.

In the case of \xsptt, we find r$_{\mathrm{plat}}=62\arcsec$, 
F$_{\mathrm{plat}}(0.5 - 2.0$~keV$) = 5.74 \pm 0.58 \times 10^{-14}$ erg s$^{-1}$ cm$^{-2}$
 and a total luminosity
L$_{\mathrm{plat}}(0.5 - 2.0$~keV$) =~2.84~\pm~0.3~\times~10^{44}$~erg~s$^{-1}$ (Fig.~\ref{fig:gca2}).

\subsubsection{X-ray spectroscopy}
\label{sec:xspec}
\begin{table*}
\begin{center}
\caption{Basic X-ray parameters of \xspt\ and \xsptt. Flux and luminosity errors include the Poisson errors and a 5\% systematic error in the background estimation. Errors of parameters obtained from scaling relations include 
the measurement errors of the luminosity and temperature, respectively, and the intrinsic scatter of the scaling relations. We assume self similar evolution for all the scaling relations and no evolution of their intrinsic scatters (see Sec.~\ref{sec:xspec}).}
\label{tab:pars}
\centering
\begin{tabular}{ l l l l}
\hline
parameter & \xspt & \xsptt & units\\
\hline
 $\alpha$ (J2000)$^a$ & \,\,\,\,$23^{\mathrm{h}}\,32^{\mathrm{m}}\,26.7^{\mathrm{s}}$ & \,\,\,\,$23^{\mathrm{h}}\,42^{\mathrm{m}}\,45.8^{\mathrm{s}}$                                             \\
 $\delta$ (J2000)$^a$ & $-53\degr\,58\arcmin\,\,\,20.4\arcsec$                        & $-54\degr\,10\arcmin\,\,\,59.2\arcsec$                                                                   \\
%%%%%%%%%%%%%%%%%%%%%%%%%%%%%%%%%%%%%%%%%%%%%%%%%%%%%%%%%%%%%%%%%%%%%%%%%%%%%
photometric redshift & $0.32^b$ & $1.08^b$ &\\

F$_{500}  $ [$0.5-2.0$ keV] & $9.52 \pm 0.51 $      & $0.58 \pm 0.06 $          &       $10^{-13}$          erg cm$^{-2}$ s$^{-1}$ \\
L$_{500}  $ [$0.5-2.0$ keV] & $2.71 \pm 0.15 $      & $2.86 \pm 0.29 $           &      $ 10^{44}$           erg s$^{-1}$           \\
T$_{500}  $                 & $9.3 \pm 2.6 ^c$         & $4.5 \pm 1.3^f$             &       $        $          keV                    \\
r$_{500}  $                 & $1.3 \pm 0.2 ^d$         & $0.6 \pm 0.1^d$             &       $        $          Mpc                     \\
M$_{500}  $                 & $8.8 \pm 3.8 ^d  $       & $1.9 \pm 0.8^d  $             &    $ 10^{14}$             M$_{\sun}$             \\
Y$_{\mathrm{X},500}$                 & $11.6 \pm 9.7^e  $       & $1.1 \pm 0.7^g  $             &    $ 10^{14}$             M$_{\sun}$ keV         \\
r$_{200}  $                 & $2.0 \pm 0.3 ^d$         & $0.9\pm0.1^d$             &       $        $          Mpc                     \\
M$_{200}  $                 & $12.4 \pm 5.4 ^d $       & $2.7 \pm 1.2^d  $             &    $ 10^{14}$             M$_{\sun}$             \\
\hline
\end{tabular}
\end{center}
\footnotesize{$^a$~X-ray coordinates based on a maximum likelihood fit of a PSF folded
 beta model to the surface brightness distribution;
$^b$~\citet{high10};
$^c$~Spectroscopic temperature (Sec.~\ref{sec:xspec}), error bars averaged, assuming isothermality;
$^d$~M$-$T relation from \citet{arnaud05}, using relations for $T~>~3.5$~keV,
self-similar evolution, radii calculated analytically from the mass estimates;
$^e$~M$_{500}-$Y$_{\mathrm{X}}$ relation \citep{arnaud07}.
$^f$~L$-$T relation from \citet{pratt09}, self-similar evolution, relation for the $0.5-2$~keV
 luminosity, BCES orthogonal fit.
$^g$ L$-$Y$_{\mathrm{X}}$ relation from \citet{pratt09}, self-similar evolution,
relation for the $0.5-2$~keV luminosity, BCES orthogonal fit. Y$_{X}$ , the X-ray analogue
 to the Comptonization parameter Y, is the product of the gas mass and temperature
 \citep[e.g.][]{kravtsov06};
}
\end{table*}
The available survey data\footnote{Due to current limitations of the \texttt{backscale} task, used to
calculate the area scaling factors of the spectra, we omit
the slew part of the survey for spectroscopical purposes and filter out from
the mosaic eventlist only events detected during the relevant \emph{stable
pointing} period (the slew part would contribute only few tens of counts in this case).}, although modest in exposure, allow us to get a first
temperature estimate for \xspt.

In order to determine a suitable aperture
for spectroscopic measurements, we create a wavelet reconstruction
\citep[][]{vikhlinin_wavelet} of the combined 0.5-2.0 keV band image.
We find that a circular aperture with 70" radius well encloses the region where
the cluster emission is registered at $\geq 5 \sigma$ significance.

A background spectrum is extracted
from an annulus concentric with the source and spanning the radial distance from $200\arcsec$ to $400\arcsec$.
The inner radius is selected based on the growth curve analysis as the radius
where cluster emission is no longer observable (Fig.~\ref{fig:gca1}). The outer
radius is constrained by the field of view. We excise all detected point sources
from each extracted spectrum after a visual check.

We fit the spectrum with a single temperature MEKAL model, fixing the column
density to the galactic value $n_{\mathrm{H}} = 1.62 \times 10^{20}$ cm$^{-2}$ \citep{dl},
metal abundance to Z=0.3 Z$_{\sun}$ and redshift to z=0.32 \citep[photometric,][]{high10}.
 To avoid biases stemming from
analysing low-count spectra, we use a minimally binned spectrum ($\ge 1$ cts/bin)
and C-statistics. 

The fitted temperature is T $=9.3^{+3.3}_{-1.9}$~keV
%!val!%
($1 \sigma$ errors) for the joint fit from all three cameras
(Fig.~\ref{fig:xspec1}). 
In order to check for possible systematics in the background subtraction, we also fit
 the spectrum using background spectra extracted from a
completely independent circular region (%$\alpha=353.3982, \delta=-53.8933$, radius=255",
on different chips than the source but roughly at the same
off-axis angle). The test background gives a consistent result,
%!val!%
T $=9.4^{+3.5}_{-1.9}$~keV.

Based on our temperature measurement we estimate several important physical
 parameters (Tab.~\ref{tab:pars}), including mass in r$_{500}$ and r$_{200}$ apertures
from the M$-$T scaling relation,
assuming self-similar evolution. For parameters obtained from scaling relations we include 
the measurement errors of the luminosity and temperature and the intrinsic scatter of the scaling relations. We use a beta model to extrapolate
the observed flux and luminosity out to r$_{500}$ (since the estimated r$_{500}$ value is larger than the measured r$_{\mathrm{plat}}$).
This extrapolation is negligible ($\sim1.5\%$).

The available photon statistics for \xsptt\ is much lower and allows us to carry out only a tentative analysis. Following the previously described procedure,
 we extract the source spectrum from a $45\arcsec$ region and background spectrum from a concentric annulus with
$100\arcsec$ inner and $200\arcsec$ outer radius. Fixing the column
density to the galactic value $n_{\mathrm{H}} = 1.86 \times 10^{20}$ cm$^{-2}$,
metal abundance to Z=0.3 Z$_{\sun}$ and redshift to z=1.08 \citep[photometric,][]{high10},
we find that the spectrum is consistent with a single temperature MEKAL model (Fig.~\ref{fig:xspec2}).
 The temperature is only weakly constrained, T $=6.7^{+5.2}_{-2.4}$~keV,
 and therefore we opt for the use of luminosity based scaling relations (L$-$T and L$-$Y$_{\mathrm{X}}$)
to estimate the physical parameters of the system (Tab.~\ref{tab:pars}). 
The evolution of the scaling relations and their intrinsic scatter is currently not firmly established out to z~1.
We assume self-similar evolution of the scaling relations and no evolution of their intrinsic scatters.
The error from these assumptions for \xsptt\ is expected to be smaller than the quoted measurement errors.

\section{Discussion and conclusions}
\label{sec:discuss}
We have presented first results from the XMM-BCS cluster survey, providing X-ray
detections of two SZE selected systems, \xspt\ and \xsptt.
The X-ray analysis is based on mosaic mode \xmm\ observations - the first time observations
 of this kind have been carried out.

\xspt\ ranks among the hottest known clusters (T $=9.3$~keV) and is exceptionally massive
 (M$_{200} \gtrsim 1 \times 10^{15}$~M$_{\sun}$).
V10 reports a SZE inferred mass from SPT of
M$_{500}$ = 5.20 $\pm$ 0.86 $\pm$ 0.83 $\times$ 10$^{14}$~M$_{\odot}$,
where the error bars represent the statistical and systematic uncertainties, both at
68\% confidence.  However, they note that this mass is biased low by the
presence of a bright dusty point source identified in the 220 GHz SPT
data.  Preliminary analysis indicates that this point source decreases the
SZE mass estimate by a factor of $\sim$1.5.  This would imply a corrected SZE mass estimate of
M$_{500}$ = 7.8 $\pm$ 1.3 $\pm$ 1.3 $\times$ 10$^{14}$~M$_{\odot}$, in good agreement
with our X-ray estimate.
This is an initial study of the system \xspt. Other aspects of the system are being investigated in forthcoming papers:
\textbf{1)} using a deeper XMM observation, Andersson et al. (in prep.), do a more
detailed comparison of the X-ray and SZ properties of this cluster, \textbf{2)} characterization of the galaxy population and morphology study will be addressed in Song et al. (in prep.).

\xsptt\ belongs to the one of the most distant known clusters (z=1.08) with X-ray and SZE detections. The discovery of such a distant system in
 both SZE and X-ray surveys demonstrates the great potential of the two observational approaches for cosmological and cluster evolution studies.
The estimated mass for this system,  M$_{500} = 1.9 \pm 0.8 \times 10^{14}$~M$_{\sun}$, is consistent
with the SZE mass M$_{500}$ = 2.66 $\pm$ 0.50 $\pm$ 0.37 $\times$ 10$^{14}$~M$_{\odot}$ (V10).

\xspt\ and \xsptt\ are the first galaxy clusters
discovered independently in X-ray, SZE and optical surveys. These
clusters showcase the promise of multi-wavelength cluster surveys and give
a glimpse of the possible synergies of current and future large-scale
survey experiments, including SPT, \emph{Planck}, \emph{eRosita}, and the Dark Energy
Survey.

\begin{acknowledgements}
We thank the Blanco Cosmology Survey team for executing and processing the observations used
in this paper. We thank the SPT team for SZE mass estimates and maps of this cluster. The South Pole Telescope is supported by the National Science Foundation
through grants ANT-0638937 and ANT-0130612.  The SPT also thanks the
National Science Foundation (NSF) Office of Polar Programs, the United
States Antarctic Program and the Raytheon Polar Services Company for their
support of the project. We also thank the \emph{XMM-Newton} SOC for implementing the mosaic mode without which
these observations couldn't have been carried out. Particularly, we thank Pedro
Rodriguez for useful discussions about the data analysis. RS acknowledges support by the DfG 
in the program SPP1177. HB acknowledges support for the research group through The Cluster of Excellence 'Origin and Structure of the Universe',
funded by the Excellence Initiative of the Federal Government of Germany, EXC project number 153.
 BB acknowledges additional support from a KICP Fellowship.
\end{acknowledgements}

\bibliographystyle{aa}
\bibliography{clusters}

\Online
\begin{appendix}
\section{Supplementary electronic material}
The Appendix provides additional information concerning the X-ray analysis \xspt\ and \xsptt.
\begin{figure}[h!]
\begin{center}
\includegraphics[width=0.5\textwidth]{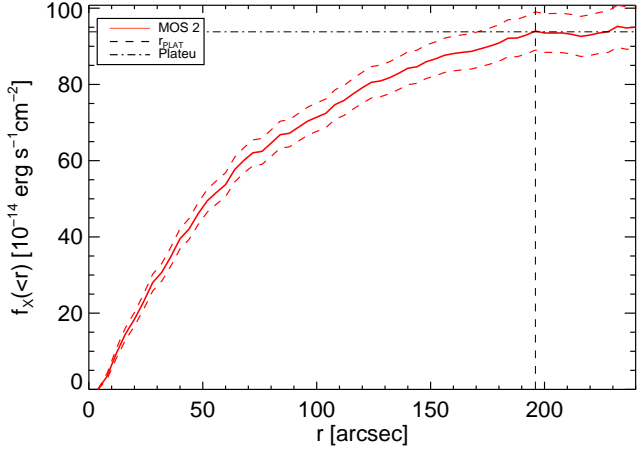}
\end{center}
\caption{The growth curve of \xspt: red curve shows the integrated flux as a function of outer integration radius for the MOS2
camera in the $0.5-2.0$~keV band. Error bars on the growth curve indicate the Poisson error of the flux measurement including a 5\% systematic
 error in the background estimation. Source flux is detected out to $196\arcsec$ (r$_{\mathrm{plat}}$, dashed line).
The total measured flux in this aperture and band is
F$_{\mathrm{plat}} = 9.4 \pm 0.5 \times 10^{-13}$ erg s$^{-1}$ cm$^{-2}$ (horizontal dot-dashed level).
Since the estimated r$_{500}$ value ($281\arcsec$) is larger than the measured r$_{\mathrm{plat}}$,  we use a beta model to
extrapolate the observed flux and luminosity out to r$_{500}$. The needed
extrapolation is however negligible ($\sim1.5\%$). 
We omit the use of the pn camera for the
growth curve analysis because of the significant losses due to
flaring and the use of the MOS1 camera because of the missing data on the lost MOS1 CCD\#6.}
\label{fig:gca1}
\end{figure}

\begin{figure}[t!]
\begin{center}
\includegraphics[width=0.5\textwidth]{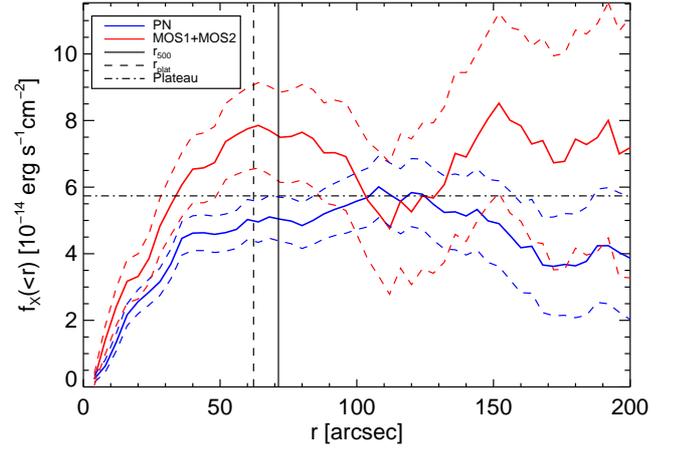}
\end{center}
\caption{The growth curve of \xsptt: red curve shows the integrated flux as a function of 
outer integration radius for the joint MOS1 and MOS2
cameras ($0.5-2.0$~keV band), blue curve for the pn camera. Error bars on the growth curve indicate the Poisson error of the flux measurement including a 5\% systematic
 error in the background estimation. Source flux is detected out to r$_{\mathrm{plat}}=62\arcsec$ (dashed line).
The total measured flux, estimated as the weighted 
average of the MOS and pn plateau fluxes, is F$_{\mathrm{plat}} = 5.7 \pm 0.6 \times 10^{-14}$ erg s$^{-1}$ cm$^{-2}$ (dot-dashed level).
Since the estimated r$_{500}$ value ($71\arcsec$, solid line)
is larger than the measured r$_{\mathrm{plat}}$,  we used a beta model to
extrapolate the observed flux and luminosity out to r$_{500}$. The needed
extrapolation is negligible ($\sim0.8\%$).}
\label{fig:gca2}
\end{figure}

\clearpage

\begin{figure}[ht!]
\begin{center}
\includegraphics[angle=-90,width=0.5\textwidth]{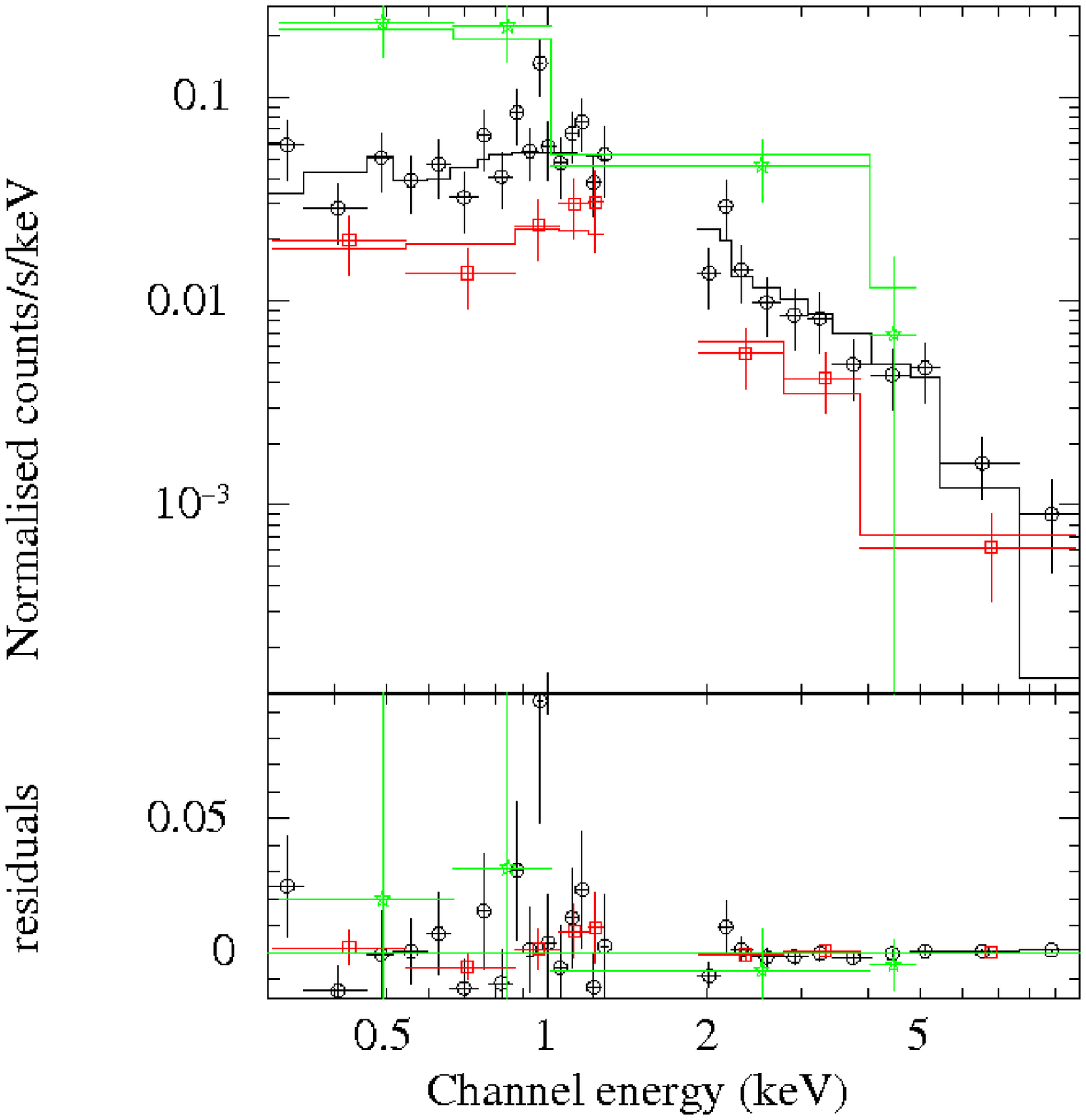}
\end{center}
\caption{\xmm\ X-ray spectrum of \xspt\ 
fitted with a single temperature MEKAL model which gives T $=9.3^{+3.3}_{-1.9}$~keV. The column density is fixed to the galactic value $n_{\mathrm{H}} = 1.62 \times 10^{20}$ cm$^{-2}$ \citep{dl}, the metal abundance to Z $=0.3$ Z$_{\sun}$ and the redshift to z $=0.32$ (photometric). The spectra were binned only
for display purposes, the fit was carried out with $\ge 1$ cts/bin binning and C-statistic.
 Red: MOS1, black: MOS2, green: pn.  See Sec.~\ref{sec:xspec} for details.}
\label{fig:xspec1}
\end{figure}

\begin{figure}[ht!]
\begin{center}
\includegraphics[angle=-90, width=0.5\textwidth]{xmmspt0002_001_nice.ps}
\end{center}
\caption{\xmm\ X-ray spectrum of \xsptt. The available low count spectrum (displayed for completeness) allows us to draw only tentative conclusions. 
The spectrum is consistent with a single temperature MEKAL model, with temperature only weak constrained to T $=6.7^{+5.2}_{-2.4}$~keV. 
The column density is fixed to the galactic value $n_{\mathrm{H}} = 1.86 \times 10^{20}$ cm$^{-2}$ \citep{dl}, 
the metal abundance to Z $=0.3$ Z$_{\sun}$ and the redshift to z $=1.08$ (photometric). The spectra were binned 
only for display purposes, the fit was carried out with $\ge 1$ cts/bin binning and C-statistic.
 Red: MOS1, black: MOS2, green: pn.  See Sec.~\ref{sec:xspec} for details.}
\label{fig:xspec2}
\end{figure}

\clearpage

\begin{figure*}[]
	\begin{minipage}{0.5\textwidth}
	\begin{center}
		\includegraphics[scale=0.49]{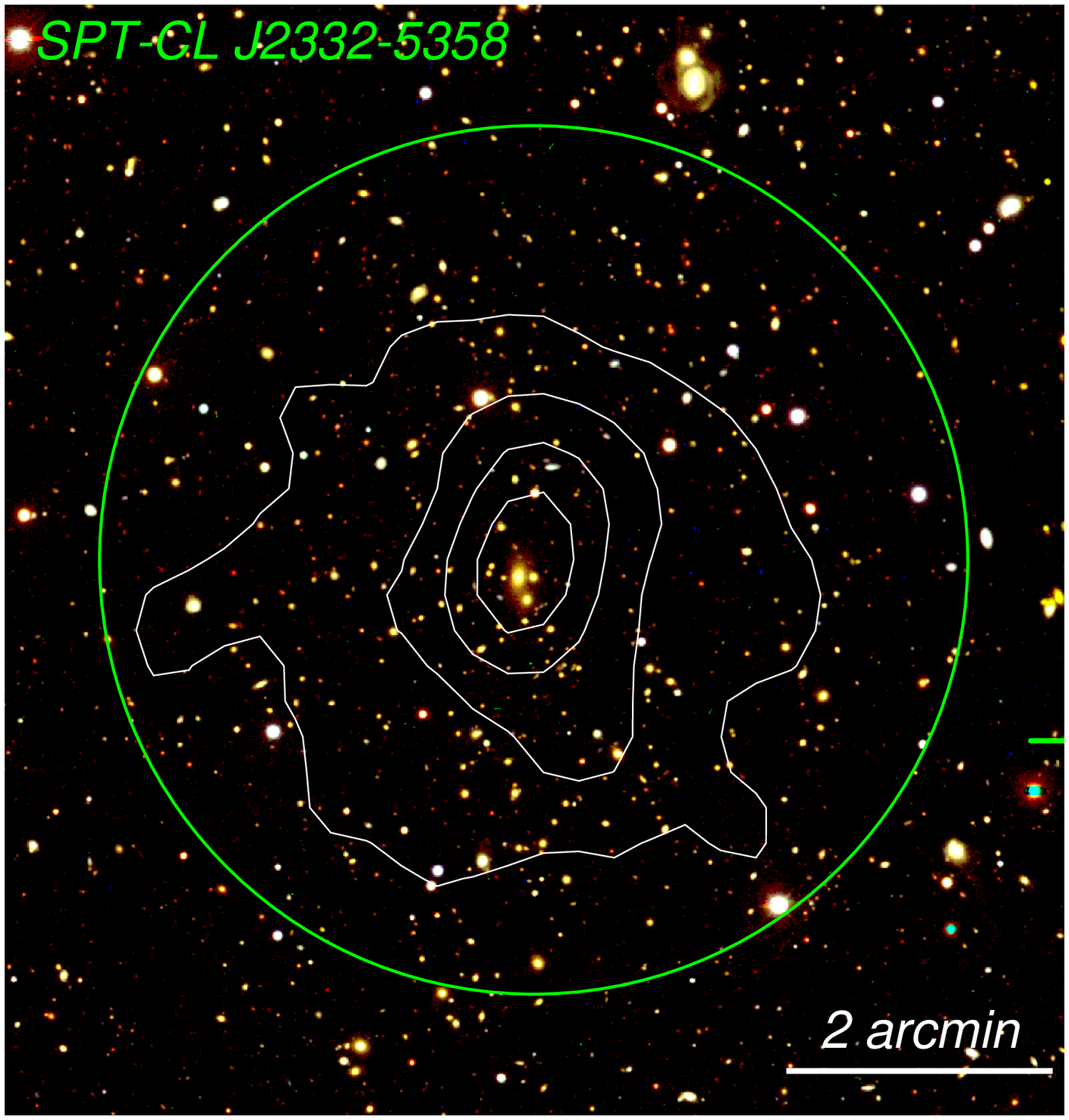}
	\end{center}
	\end{minipage}
	\begin{minipage}{0.5\textwidth}
	\begin{center}
		\includegraphics[scale=0.49]{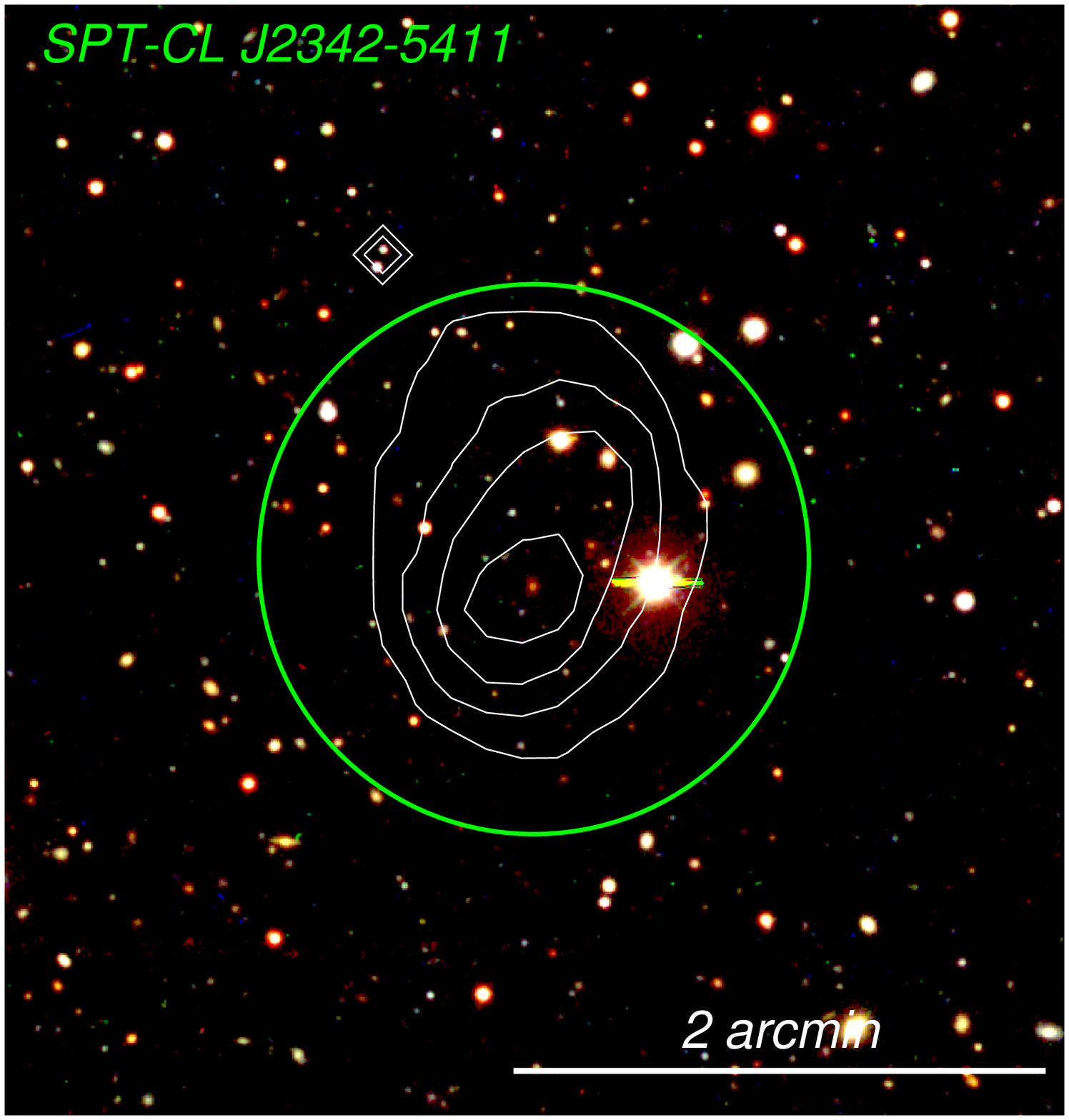}
	\end{center}
	\end{minipage}
\caption{\emph{Left: } Color image of \xspt.
\emph{Right: } Color image of \xsptt.
Both images were obtained from the Blanco Cosmology Survey imaging in the $gri$ bands.
X-ray contours are overlaid in white. Green circles show the estimated r$_{\mathrm{plat}}$ (see Sec.~\ref{sec:gca}).
Both clusters have a large BCG (brightest cluster galaxy) within few arcseconds ($\lesssim 10\arcsec$) from the
X-ray emission peak.}
\label{fig:opt}
\end{figure*}

\end{appendix}
\end{document}